\documentclass[useAMS,usenatbib]{mn2e}
\usepackage{epsfig}
\usepackage{color}
\usepackage{deluxetable}
\usepackage{rotating}
\usepackage{amssymb}
\usepackage{graphicx}
\usepackage{subfigure}
\usepackage{journals}
\usepackage{psfrag}
\usepackage{amsmath}
\usepackage{amssymb}
\usepackage{lscape}
\usepackage{IEEEtrantools}
\def\reff@jnl#1{{\rm#1\/}}
\def\aj{\reff@jnl{AJ}}                 % Astronomical Journal
\def\araa{\reff@jnl{ARA\&A}}           % Annual Review of Astron and Astrophys
\def\apj{\reff@jnl{ApJ}}               % Astrophysical Journal
\def\apjl{\reff@jnl{ApJ}}              % Astrophysical Journal, Letters
\def\apjs{\reff@jnl{ApJS}}             % Astrophysical Journal, Supplement
\def\ao{\reff@jnl{Appl.Optics}}        % Applied Optics
\def\apss{\reff@jnl{Ap\&SS}}           % Astrophysics and Space Science
\def\aap{\reff@jnl{A\&A}}              % Astronomy and Astrophysics
\def\aapr{\reff@jnl{A\&A~Rev.}}        % Astronomy and Astrophysics Reviews
\def\aaps{\reff@jnl{A\&AS}}            % Astronomy and Astrophysics, Supplement
\def\azh{\reff@jnl{AZh}}               % Astronomicheskii Zhurnal
\def\baas{\reff@jnl{BAAS}}             % Bulletin of the AAS
\def\jrasc{\reff@jnl{JRASC}}           % Journal of the RAS of Canada
\def\memras{\reff@jnl{MmRAS}}          % Memoirs of the RAS
\def\mnras{\reff@jnl{MNRAS}}           % Monthly Notices of the RAS
\def\pra{\reff@jnl{Phys.Rev.A}}        % Physical Review A: General Physics
\def\prb{\reff@jnl{Phys.Rev.B}}        % Physical Review B: Solid State
\def\prc{\reff@jnl{Phys.Rev.C}}        % Physical Review C
\def\prd{\reff@jnl{Phys.Rev.D}}        % Physical Review D
\def\prl{\reff@jnl{Phys.Rev.Lett}}     % Physical Review Letters
\def\pasp{\reff@jnl{PASP}}             % Publications of the ASP
\def\pasj{\reff@jnl{PASJ}}             % Publications of the ASJ
\def\qjras{\reff@jnl{QJRAS}}           % Quarterly Journal of the RAS
\def\skytel{\reff@jnl{S\&T}}           % Sky and Telescope
\def\solphys{\reff@jnl{Solar~Phys.}}   % Solar Physics
\def\sovast{\reff@jnl{Soviet~Ast.}}    % Soviet Astronomy
\def\ssr{\reff@jnl{Space~Sci.Rev.}}    % Space Science Reviews
\def\zap{\reff@jnl{ZAp}}               % Zeitschrift fuer Astrophysik
\def\nat{\reff@jnl{Nature}}            % Nature
\title[Weak emission line galaxies] {Radio continuum detection in blue early-type weak emission line galaxies} \author[A. Paswan and
  A. Omar] {A.~Paswan$^{1,2}$\thanks{E-mail: p.abhishek@aries.res.in},
  and A.~Omar$^{1,2}$\\$^{1}$Aryabhatta Research Institute of Observational SciencES, Manora Peak, Nainital 263002, India\\$^{2}$Pt. Ravishankar Shukla University, Raipur, 492010, India}

\date{Accepted ---. Received ---; in original form \today}
\pagerange{\pageref{firstpage}--\pageref{lastpage}}
\pubyear{2011}

\begin{document}
\label{firstpage}
\maketitle

\begin{abstract}

The star formation rates (SFRs) in weak emission line (WEL) galaxies in a volume-limited ($0.02 < z < 0.05$) sample of blue early-type galaxies (ETGs) identified from SDSS, are constrained here using 1.4 GHz radio continuum emission. The direct detection of 1.4 GHz radio continuum emission is made in 8 WEL galaxies and a median stacking is performed on 57 WEL galaxies using VLA FIRST images. The median stacked 1.4 GHz flux density and luminosity are estimated as 79 $\pm$ 19 $\mu$Jy and 0.20 $\pm$ 0.05 $\times$ 10$^{21}$ W Hz$^{-1}$ respectively. The radio far-infrared correlation in 4 WEL galaxies suggests that the radio continuum emission from WEL galaxies is most likely due to star formation activities. The median SFR for WEL galaxies is estimated as 0.23 $\pm$ 0.06 M$_{\odot}$yr$^{-1}$, which is much less compared to SFRs ($0.5 - 50$ M$_{\odot}$yr$^{-1}$) in purely star forming blue ETGs. The SFRs in blue ETGs are found to be correlated with their stellar velocity dispersions ($\sigma$) and decreasing gradually beyond $\sigma$ of $\sim 100$ km s$^{-1}$. This effect is most likely linked with the growth of black hole and suppression of star formation via AGN feedback. The color differences between SF and WEL sub-types of blue ETGs appear to be driven to large extent by the level of current star formation activities. In a likely scenario of an evolutionary sequence between sub-types, the observed color distribution in blue ETGs can be explained best in terms of fast evolution through AGN feedback. 
\end{abstract}

\begin{keywords}
galaxies: evolution -- radio continuum: galaxies -- techniques: image processing  -- methods: statistical
\end{keywords}

%%%%%%%%%%%%%%%%%%%%%%%%%%%%%%%%%%%%%%%%%%%%%%%%%%%%%%%%%
\section{Introduction} %\label{sec:intro}
%%%%%%%%%%%%%%%%%%%%%%%%%%%%%%%%%%%%%%%%%%%%%%%%%%%%%%%%%

Early-type galaxies (ETGs) were once believed to be a single type of objects passively evolving without any presence of cold interstellar medium (ISM) and star formation. The $Infrared ~Astronomical ~Satellite ~(IRAS)$ detected for the first time a presence of cold ISM in the form of dust in elliptical galaxies in the local Universe \citep{1984PASP...96..973N,1985AJ.....90..454K,1989ApJS...70..329K}. The cold ISM was subsequently detected in several elliptical galaxies through HI observations \citep{1989AJ.....97..708V, 1995A&A...300..675H}. \citet{2006ApJ...644..850S} detected HI in a volume-limited sample of S0 galaxies. \citet{2006MNRAS.371..157M} detected HI along with radio continuum in 12 S0 galaxies in which ionized gas in the centers was already known to exist. The molecular contents responsible for star formation in elliptical and S0 galaxies have been detected through CO and OH observations of several infrared bright elliptical galaxies \citep{1992AJ....104.2097W,1995A&A...297..643W,1996ApJ...460..271K,2002A&A...381L..29O, 2007MNRAS.377.1795C}. A sample of ETGs having star formation showed presence of ionized gas traced by H$_{\alpha}$ \citep{1996A&AS..120..463M}. It is well established that ETGs belong to a class of heterogeneous objects with varying physical conditions and compositions in the ISM, and encompass a large range in star formation rate \citep{1992ApJ...387..484B, 1993AJ....106..907H, 1999A&AS..136..269F, 2000AJ....120..165T, 2006MNRAS.371..157M, 2007MNRAS.382.1415S, 2009MNRAS.396..818S, Huang2009MNRAS.398.1651H, 2014MNRAS.442..533M}. 

The galaxy zoo project using the $Sloan ~Digital ~Sky ~Survey ~(SDSS)$ data has resulted into detection of several  ETGs in the nearby Universe with a large scatter in their optical colors and emission line strengths \citep{2003RMxAC..17..167B, 2004ApJ...601L.127F, 2007MNRAS.382.1415S, 2014MNRAS.440..889S}. A new class of 'blue' ETG was inferred based on their distinct \textit{$u - r$} colors, significantly bluer than the classical red ETGs \citep[][hereafter SCH09]{2009MNRAS.396..818S, 2014MNRAS.440..889S}. The \textit{$u - r$} colors of blue ETGs are in the range of $1.3 - 2.5$, while red sequence galaxies normally have colors between $2.5 - 3.0$. The detection of several emission lines typical of star forming regions and active galactic nuclei (AGN) in star forming ETGs established that there is copious amount of cold and ionized ISM associated with star formation as well as presence of AGN related activities in these objects. \citep{2006MNRAS.371..157M, 2006MNRAS.366.1151S, 2007MNRAS.382.1415S, 2009ApJ...692L..19S, 2014MNRAS.440..889S}.

The nature of optical emission lines from galaxies is distinguished into star formation, AGN or composite using the BPT (Baldwin, Phillips \& Terlevich) diagram \citep{1981PASP...93..817B,1987ApJS...63..295V}. In the SCH09 volume-limited sample having a total of 204 blue ETGs, ~25\% galaxies are classified as purely star forming (SF), ~12\% are mainly AGN (Seyferts and low ionization emission line regions -- LINERs), ~25\% host both AGN and star formation, and rest (38\%) show no or weak emission lines. The last group of objects known as weak emission line (WEL) galaxies constitute a sample of 76 galaxies. Significant patterns were seen in color-mass diagram as well as in optical line strengths in star-forming (SF), AGN+SF composite, AGN, WEL sub-types of blue ETGs. Galaxies having strong emission lines indicative of star formation are the bluest in color and significantly off from the red sequence while WEL galaxies are located closer to the red sequence \citep{2007MNRAS.382.1415S, Huang2009MNRAS.398.1651H}. The colors of galaxies exhibiting spectrum of AGN and AGN+SF composite are found between these two extremes \citep{2014MNRAS.440..889S}. This observed pattern in average colors of different types of ETGs suggests some kind of evolutionary sequence within different sub-types of blue ETGs \citep{2004ApJ...608..752B, 2006MNRAS.368..414D, 2007MNRAS.382.1415S, 2009MNRAS.393.1324B, Huang2009MNRAS.398.1651H, 2014MNRAS.442..533M, 2014MNRAS.440..889S}. The colors also indicate that WEL sub-type has some links with the AGN activities. 

Constraining star formation rate (SFR) in different sub-types of blue ETGs can be of significant importance for understanding evolution of ETGs. The SFR has been estimated for purely star forming blue ETGs in the range $0.5 - 50$ M$_{\odot}$yr$^{-1}$ using optical emission lines in SCH09. It was not possible to constrain SFR in WEL galaxies using the optical data since the detections of emission lines are of low signal-to-noise ratio in WEL galaxies. Constraining SFR in AGN+SF will be very challenging since contribution of AGN needs to be first determined and removed. The SFR in WEL galaxies can possibly be derived from the 1.4 GHz radio continuum emission if any residual AGN contribution to the radio continuum emission is negligible. Given the weakness of optical line strengths in WEL galaxies, the SFR in these galaxies are expected to be significantly less than 1 M$_{\odot}$yr$^{-1}$ (SCH09). 

We searched for 1.4 GHz radio continuum emission in WEL galaxies in the sample of SCH09 using the Faint Images of Radio Sky at Twenty-cm (FIRST) survey images taken with the Very large Array (VLA) \citep{1995ApJ...450..559B}. We detected radio continuum emission in 8 WEL galaxies. We used image-stacking technique to obtain limits on SFR in 57 WEL galaxies. We detected significant radio continuum emission in the stacked image. This paper presents the radio detection statistics and the SFR limits to the WEL galaxies. This paper also makes a brief discussion on implications of detection of radio continuum in blue ETGs on AGN feedback process in galaxies. The relevant cosmological parameter used in this paper is H$_{o}$ = 73.0 km s$^{-1}$ Mpc$^{-1}$.

%%%%%%%%%%%%%%%%%%%%%%%%%%%%%%%%%%%%%%%%%%%%%%%%%%%%%%%%%
\section{Sample properties and radio detection in WEL galaxies} %\label{sec:data sample}
%%%%%%%%%%%%%%%%%%%%%%%%%%%%%%%%%%%%%%%%%%%%%%%%%%%%%%%%%

The blue early-type galaxy sample of SCH09 is a volume-limited sample of 204 galaxies in the redshift range of $0.02 - 0.05$. This sample lists a total of 50 star forming (SF), 52 AGN+SF composite, 26 AGN, and 76 WEL galaxies. The redshift distributions of WEL and SF galaxies are shown in Figure 1. The 1.4 GHz radio continuum emission was searched at the optical locations of all WEL galaxies using the archived FIRST images \citep{1995ApJ...450..559B}.  The FIRST images were available only for 65 out of 76 WEL galaxies in SCH09. A direct detection of radio continuum emission is made in only eight WEL galaxies using the FIRST images. Their properties and detection statistics are given in Table 1. We then stacked the FIRST images of the regions centered at the optical locations of the un-detected WEL galaxies. The stacked image shown in Figure 2 is a median combination of 57 FIRST images. The detection statistics in the stacked image is listed in the last row of Table 1. The radio flux density at the center is estimated as 79 $\mu$Jy with an rms of 19 $\mu$Jy beam$^{-1}$. The beam sizes in the individual images vary between $5" - 6"$. The median stacking was preferred over mean stacking since median combination is less sensitive to any chance appearance of highly deviating flux value in an individual image. 

\begin{figure}
\centering
\includegraphics[width=8.0cm,height=10.0cm]{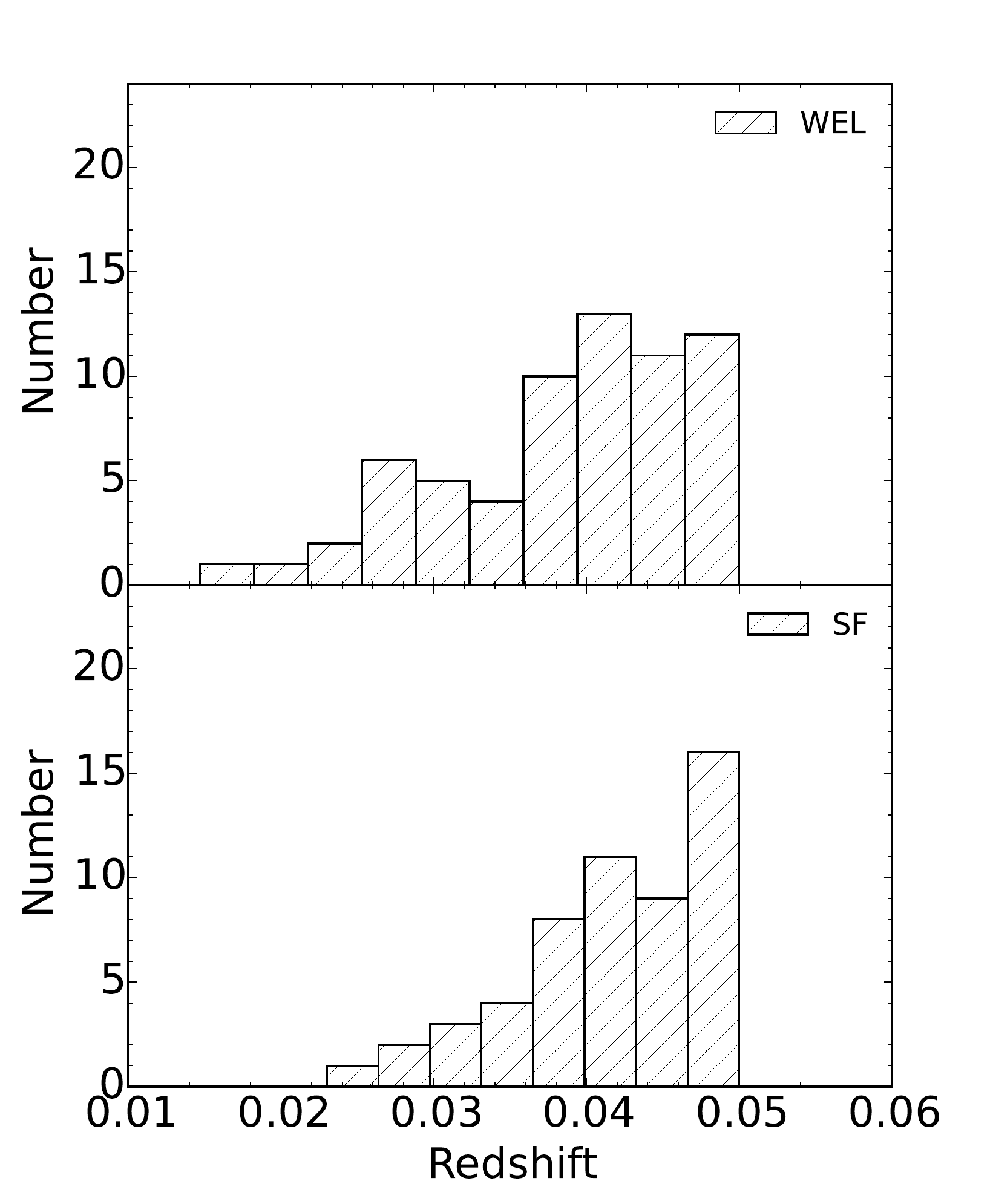}
\caption{The redshift distribution of weak emission line galaxies and purely star forming galaxies in the SCH09 sample.}
\end{figure}

\begin{figure}
\centering
\includegraphics[width=9.0cm,height=9.0cm]{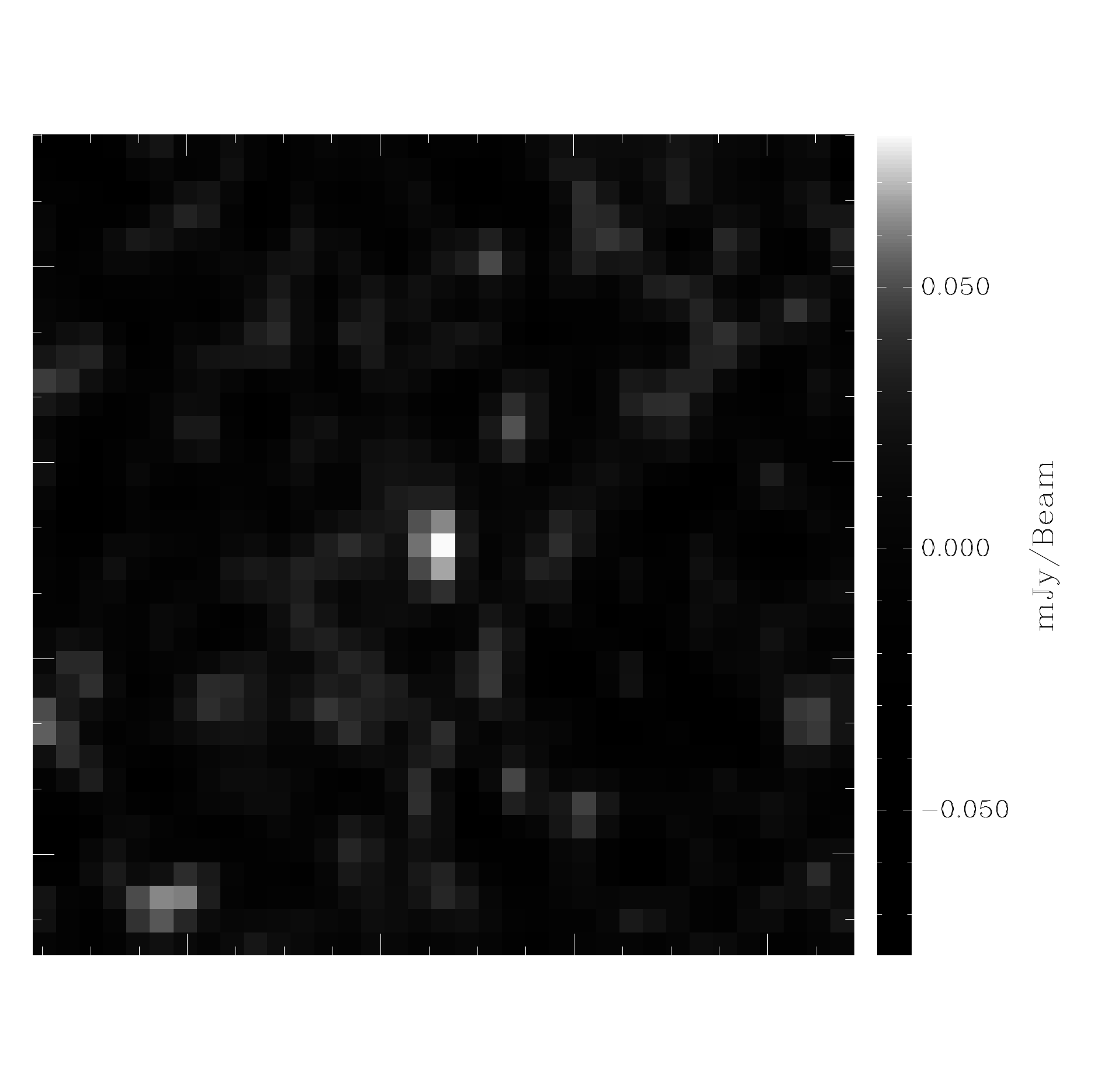}
\caption{The 1.4 GHz median stacked image of 57 weak emission line galaxies. The image size is 1' $\times$ 1' and pixel size is 1.8" $\times$ 1.8". The synthesized beam (resolution) is nearly 3 $\times$ 3 pixel in the FIRST images.}
\end{figure}

\begin{table*}
\caption{General properties of weak emission line galaxies.}
\begin{tabular}{c|c|c|c|c|c|c|c}
\hline
Galaxy & z & \textit{$u - r$} & F$_{1.4 \mathrm{GHz}}$ & L$_{1.4 \mathrm{GHz}}$ & $S_{60\mu \mathrm{m}}$ & $S_{100\mu \mathrm{m}}$ & q \\
       &   & [mag] &  [mJy]          & [W Hz$^{-1}$ $\times$ 10$^{21}$]  & [Jy] & [Jy] & \\
\hline
J003245-102703    &       0.0383$\pm$0.0002 &    2.353 &      1.12$\pm$0.14 &         3.31$\pm$0.45 & 0.14$\pm$0.04 & 0.57$\pm$0.12 & 2.45$\pm$0.07 \\      
J075853+521930    &       0.0407$\pm$0.0001 &    2.161 &      1.98$\pm$0.14 &         6.61$\pm$0.45 & 0.15$\pm$0.04 & 0.74$\pm$0.21 & 2.28$\pm$0.09 \\     
J090036+464111    &       0.0274$\pm$0.0002 &    2.000 &      1.69$\pm$0.14 &         2.57$\pm$0.25 & ----- & ----- & ----- \\              
J111733+511617    &       0.0276$\pm$0.0002 &    2.332 &      2.38$\pm$0.13 &         3.63$\pm$0.25 & 0.20$\pm$0.04 & 0.57$\pm$0.09 & 2.19$\pm$0.06 \\              
J112507+494202    &       0.0499$\pm$0.0002 &    2.382 &      1.30$\pm$0.15 &         6.45$\pm$0.79 & ----- & ----- & ----- \\              
J131349+604104    &       0.0381$\pm$0.0002 &    2.169 &      0.35$\pm$0.15 &         1.02$\pm$0.45 & 0.44$\pm$0.05 & 0.70$\pm$0.15 & 3.24$\pm$0.19 \\              
J151429+073546    &       0.0448$\pm$0.0001 &    2.042 &      0.41$\pm$0.15 &         1.66$\pm$0.61 & ----- & ----- & ----- \\              
J161325+215433    &       0.0319$\pm$0.0001 &    2.423 &      0.51$\pm$0.15 &         1.05$\pm$0.31 & ----- & ----- & ----- \\             
Stacked image  &       0.035$\pm$0.001 &    ----- &      0.08$\pm$0.02 &           0.20$\pm$0.05 & ----- & ----- & ----- \\
\hline
\end{tabular}\\
\end{table*}

The stacking of the FIRST images has been used previously in making statistically significant radio detections in K+A galaxies \citep{2012ApJ...761L..16N}, green peas galaxies \citep{2012ApJ...746L...6C} and in blue-cloud galaxies \citep{2015ApJ...803...51B}. The technique has been extensively discussed in \citet{2007ApJ...654...99W}. The detection based on image stacking relies on improving signal-to-noise ratio by a factor of $\sqrt{N}$ by stacking $N$ images in absence of any significant non-Guassian systematic noise. The rms noise estimated in the source free region of the stacked image of WEL galaxies is 19 $\mu$Jy beam$^{-1}$. This value is close to the expected $\sqrt{N}$ improvement in the signal-to-noise ratio in the FIRST images having typical rms of 0.15 mJy beam$^{-1}$. Since any kind of systematic noise in radio images can affect the detection statistics in the stacked image significantly, we checked individual images visually and through an analysis of their pixel flux distribution. The Figure 3 shows distribution of pixel flux values of 57 individual VLA FIRST images and the final stacked image. The flux distribution in all images is nearly Gaussian. This figure clearly indicates that the stacked image is not affected significantly by any systematic noise or pixel artifact. The detection made in the stacked image is therefore reliable. 

The radio luminosity is computed from the 1.4 GHz flux density using the relation given in \citet{2001ApJ...554..803Y}:

\begin{equation}\label{sfr}
\mbox{log}~\mbox{L}_{1.4GHz}[\mbox{W}~\mbox{Hz}^{-1}]~=~20.08~+~2~\mbox{log}~D~+~\mbox{log}~\mbox{S}_{1.4GHz}
\end{equation}

where $D$ represents the distance in Mpc and S$_{1.4GHz}$ represents the 1.4 GHz radio continuum flux density in Jy. Assuming a linear relationship between distance ($D$) and redshift ($z$), the equation (1) can be written as: 

\begin{equation}
\mbox{S}~\text{$\sim$}~\frac{\mbox{L}}{2 \times 10^{27}~z^{2}}
\end{equation}

In a stacking of $N$ images at different redshifts, having constant luminosity $L$, the above relation can be written as:

\begin{equation}
\langle \mbox{S} \rangle~=~\frac{1}{N}~\sum_\mathbf{i=1}^{N}\mbox{S}_{i}~=~\frac{\mbox{L}}{2 \times 10^{27}}~\frac{1}{N}~\sum_\mathbf{i=1}^{N}\frac{1}{z^{2}_{i}}~=~\frac{\mbox{L}}{2 \times 10^{27}~z^{2}_{eq}} 
\end{equation}

where,

\begin{equation}
\frac{1}{z^{2}_{eq}}~=~\frac{1}{N}~\sum_\mathbf{i=1}^{N}\frac{1}{z^{2}_{i}}
\end{equation}

where z$_{eq}$ is the equivalent redshift for obtaining a luminosity estimate directly from the detected flux in the stacked image. The z$_{eq}$ for the present sample is 0.035 $\pm$ 0.001, and the corresponding median 1.4 GHz radio luminosity for WEL galaxies is estimated as 0.20 $\pm$ 0.05 $\times$ 10$^{21}$ W Hz$^{-1}$. The radio luminosities of directly detected WEL galaxies are in the range of 1 to 7 $\times$ 10$^{21}$ W Hz$^{-1}$. 

\begin{figure*}
\includegraphics[width=7.0cm,height=8.0cm,angle=270]{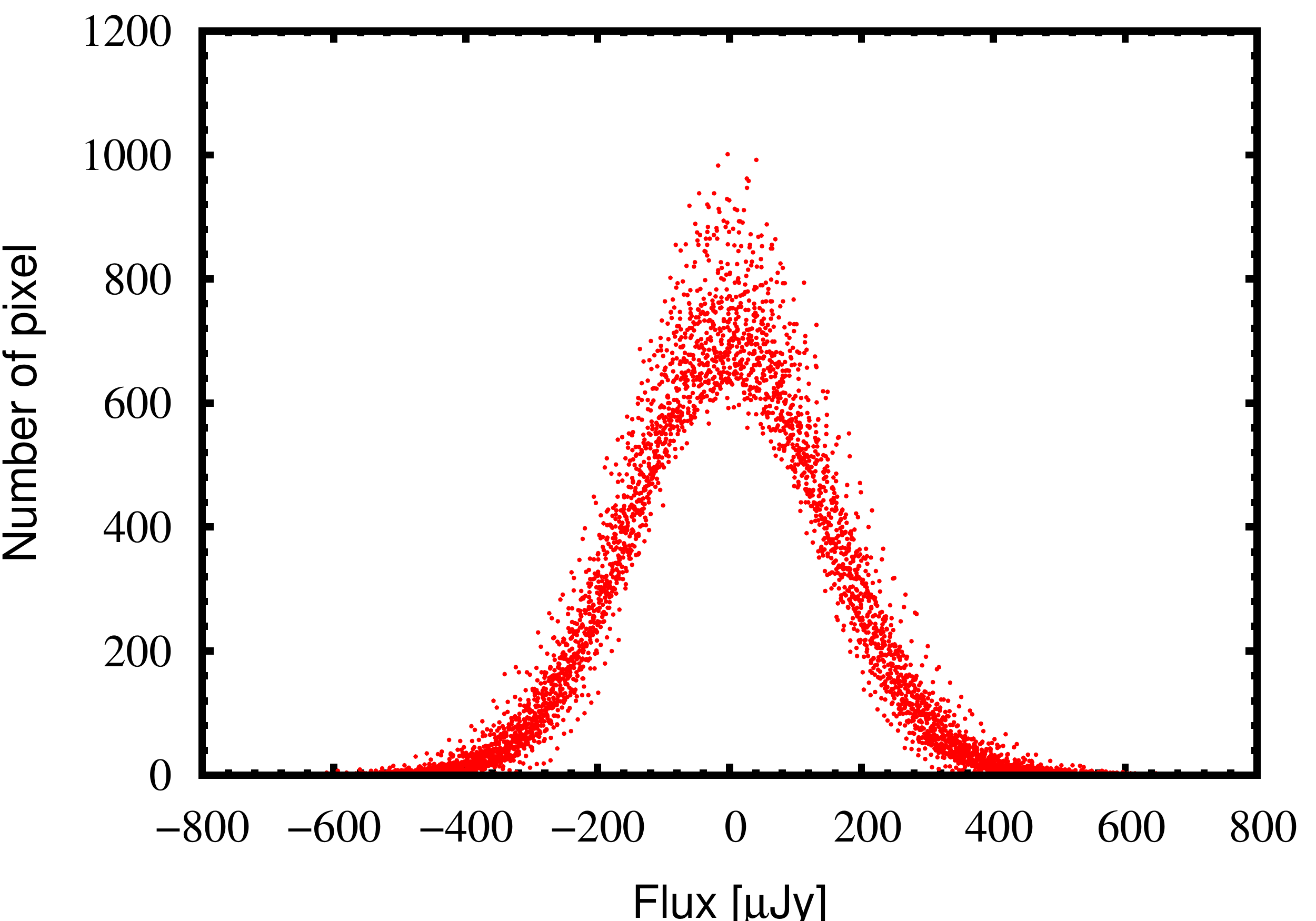}
\includegraphics[width=7.0cm,height=8.0cm,angle=270]{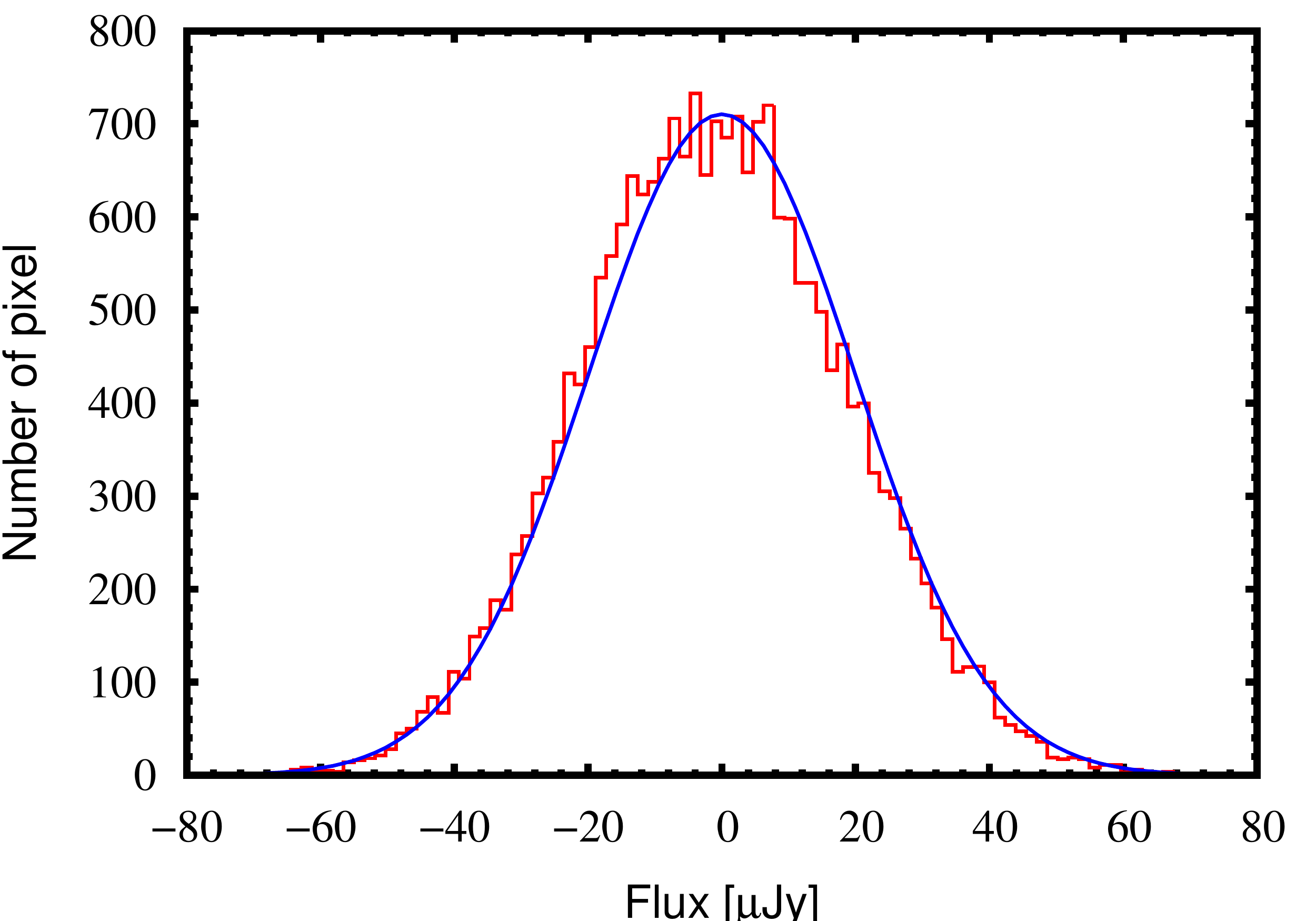}
\caption{The left panel shows pixel flux value distribution of 57 individual FIRST images used in the stacking process. The right panel shows the histogram of pixel flux value distribution for the median stacked image of 57 WEL galaxies fitted with a Gaussian.}
\end{figure*}

%%%%%%%%%%%%%%%%%%%%%%%%%%%%%%%%%%%%%%%%%%%%%%%%%%%%%%%%%
\section{Results and discussions} %\label{sec:results}
%%%%%%%%%%%%%%%%%%%%%%%%%%%%%%%%%%%%%%%%%%%%%%%%%%%%%%%%%

The radio continuum emission in nearby galaxies has both thermal and non-thermal components \citep{1992ARA&A..30..575C}. The later being due to synchrotron emission tracing its origin either in relativistic cosmic electrons accelerated in supernova process indicative of star formation activities or in central black hole related AGN activities. The thermal emission is due to star formation activities, but, is relatively very week compared to the synchrotron emission due to star formation activities \citep{1992ARA&A..30..575C}. The separation of two different synchrotron emissions (SF and AGN) can be difficult. In the SCH09 sample, galaxies are identified as AGN or SF based on an analysis of optical emission lines. Since optical lines are of poor strengths in WEL galaxies, the nature of WEL galaxies remained largely un-identified. Therefore, it is important to identify nature of radio emission from WEL galaxies from an independent analysis such as radio far-infrared correlation \citep{1973A&A....29..263V, 1992ARA&A..30..575C, 2001ApJ...554..803Y}, which can reveal any AGN related emission as an excess to that expected from the star forming activities alone. We searched for far-infrared (FIR) emission, another indicator of star formation, in eight directly detected WEL galaxies at 1.4 GHz radio continuum. The 1-D co-add IRAS scans at source locations provided by the online IRAS scan processing and integration tool (SCANPI) were used. The peak flux densities at 60$\mu \mathrm{m}$ and 100$\mu \mathrm{m}$ at source locations in the noise weighted IRAS co-added scans are given in Table 1. Four out of eight WEL galaxies have significant detections at 60$\mu \mathrm{m}$ and 100$\mu \mathrm{m}$ wavelengths. The `q' parameter as a ratio of the FIR luminosity and 1.4 GHz radio luminosity \citep{2001ApJ...554..803Y} is also computed and given in Table 1. The `q' parameter is a strong indicator of ongoing star formation activities in galaxies if value of `q' for a galaxy is between $1.6 - 3.0$, as evidenced from a study of large sample of star forming galaxies \citep{2001ApJ...554..803Y}. If value of `q' is significantly below this range, it is normally taken as an indicator of AGN related excess radio continuum emission \citep[e.g.,][]{2005JApA...26...89O}. It can be seen that the `q' values in 3 WEL galaxies are within the normal range for a star forming galaxy and one WEL galaxy shows slightly higher `q' value. No WEL galaxy shows a radio excess indicative of AGN related activities. Although our sample size is very small, it can be safely assumed that WEL galaxies do not have any significant AGN activities and the 1.4 GHz radio continuum emission is due to star formation activities. 

The radio luminosity can be converted into SFR (M $\textgreater$ 0.1 M$_{\odot}$) using the relation \citep{2002AJ....124..675C} 

\begin{equation} \label{sfr}
\mbox{SFR}~[\mbox{M}_\odot~\mbox{yr}^{-1}]~=~1.2 \times 10^{-21}~\mbox{L}_{1.4GHz}[\mbox{W}~\mbox{Hz}^{-1}]
\end{equation}

This relationship assumes a Salpeter initial mass function between $0.1 - 100$ M$_{\odot}$. This relation is obtained after subtracting a thermal emission contribution to the total emission at 1.4 GHz. The FIRST images have typical rms of $\sim$ 0.15 mJy beam$^{-1}$, which corresponds to 3$\sigma$ detection limit to SFR as 0.44 M$_{\odot}$yr$^{-1}$ and 2.74 M$_{\odot}$yr$^{-1}$ at the redshifts of 0.02 and 0.05 respectively.  The SFRs for the directly detected 8 WEL galaxies are in the range of 1 to 8 M$_{\odot}$yr$^{-1}$. The median SFR for the stacked WEL galaxies is estimated as 0.23 $\pm$ 0.06 M$_{\odot}$yr$^{-1}$. \citet{2012ApJ...755..105S} estimated SFR as $\sim0.5$ M$_{\odot}$yr$^{-1}$ in a sample of  ETGs at $z\sim0.1$, in close agreement with the median SFR derived here. 

\begin{figure}
\includegraphics[width=7.5cm,height=9.0cm,angle=270]{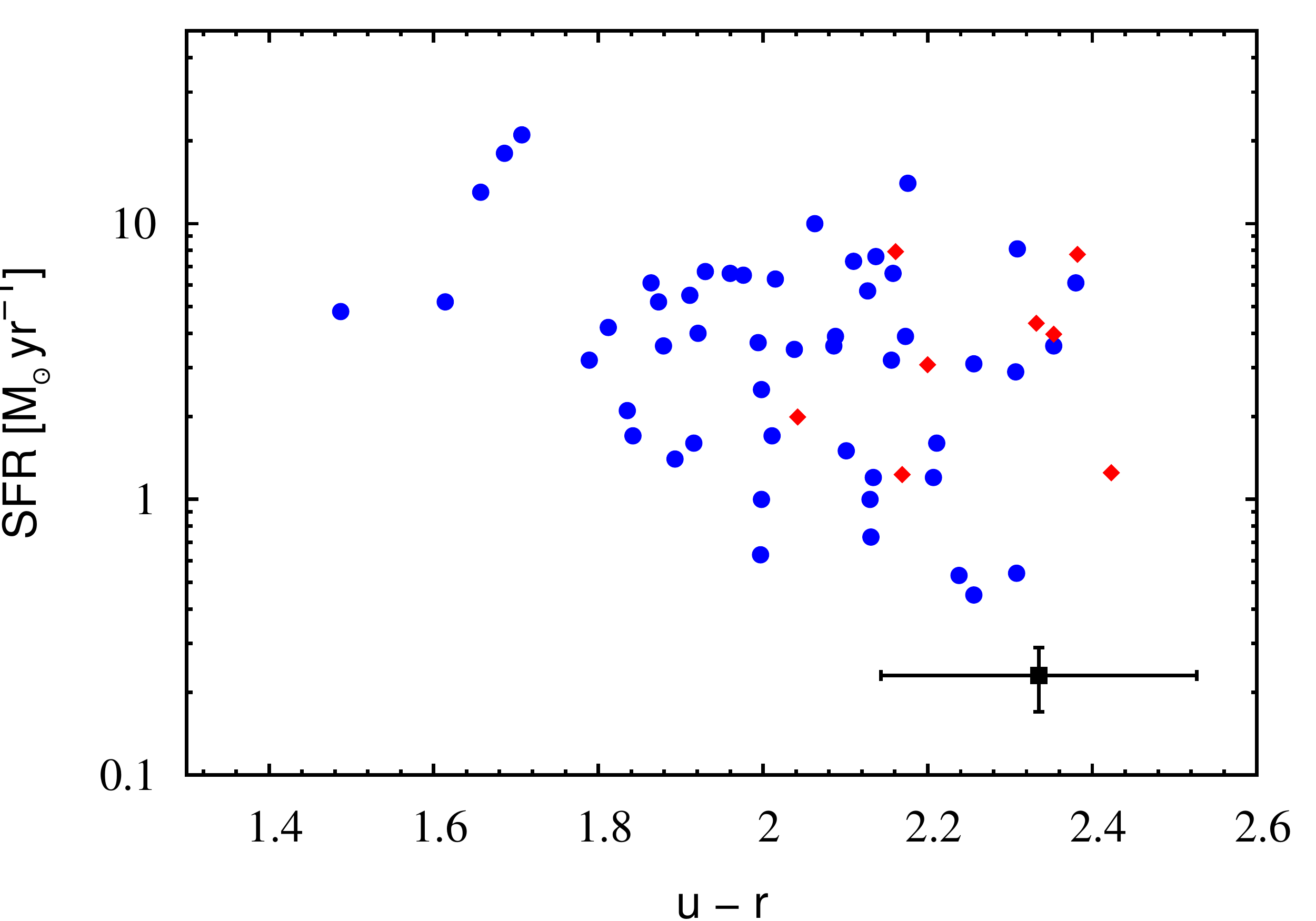}
\caption{The SFRs of blue ETGs are plotted against the $u - r$ color. The circular points are for SF-type and the diamond points are for WEL-type (directly detected in the FIRST images). The estimated median SFR for WEL galaxies using stacking of the FIRST images is shown by a square point in which the error-bar along the color axis represents range of $u-r$ color seen in WEL galaxies.}
\end{figure}

\begin{figure}
\centering
\includegraphics[width=9.5cm,height=8.0cm]{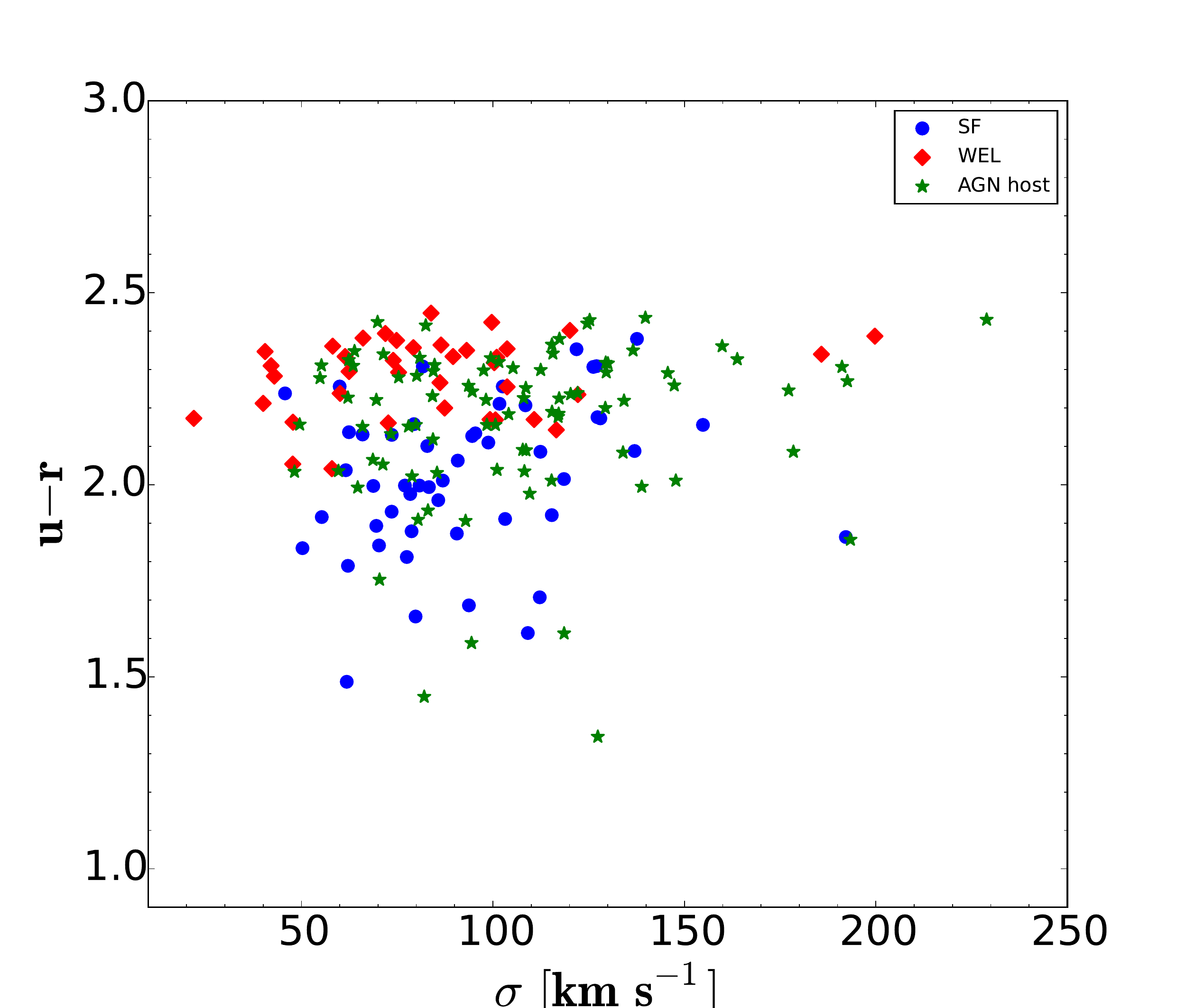}
\caption{The color-$\sigma$ relation for all sub-types (SF, AGN host, WEL) of blue ETGs in the present sample.}
\end{figure}

The ETGs have blue colors due to ongoing star formation activities. The star forming galaxies showing strong emission lines are the bluest in color with average $u-r$ color as 2.02 $\pm$ 0.02. The WEL galaxies are significantly less blue in color with average $u-r$ color as 2.31 $\pm$ 0.01. The WEL galaxies are located closer to the classical red sequence of ETGs \citep{2007MNRAS.382.1415S, Huang2009MNRAS.398.1651H}. The average color of blue ETGs with AGN activities is 2.16 $\pm$ 0.02, which is in between average colors of SF and WEL galaxies. The SFRs of blue ETGs studied here are plotted against their $u-r$ colors in Figure 4. The estimated median SFR for the WEL galaxies using stacking of the FIRST images is also shown in this figure. It can be seen that SFR values have large scatter with respect to the $u-r$ color. Nevertheless, galaxies having high SFR appear to be bluer on average than galaxies having low SFR. It can also be seen that the median value of SFR in WEL galaxies is consistent with their colors. The large scatter in SFR-color relation is most likely due to several factors coming from uncertainties in the past history of star formation episodes, dust content, metal enrichment and SFR estimates.  

The primary mechanism that regulates star formation and color evolution in ETGs is still a matter of debate \citep{2012ApJ...755..105S, 2014MNRAS.440..889S}. The stellar velocity dispersion in ETGs plays an important role in regulating star formation. The stellar velocity dispersion ($\sigma$), indicative of dynamical age and stellar mass, in blue ETGs is found to be significantly lower than the majority of red ETGs with similar luminosities \citep{2006Natur.442..888S, 2007MNRAS.382.1415S, 2009MNRAS.396..818S, Huang2009MNRAS.398.1651H, 2009MNRAS.398.2028J}. This indicates that blue ETGs are dynamically younger and less massive compared to their red-sequence counterparts. It was also observed that fraction of blue ETGs drops significantly at the high end ($\sim 200$ km s$^{-1}$) of their stellar velocity dispersion distribution and is largely independent of the optical luminosity \citep{2006Natur.442..888S}. The stellar velocity dispersion of blue ETGs of SCH09 are plotted in Figure 5 for all sub-types (SF, AGN host, WEL). It can be seen in this figure that there is no appreciable difference between their sub-types in term of $\sigma$. It can also be seen that blue ETGs are found mainly with $\sigma$ $\textless$ 150 km s$^{-1}$. 

In order to understand a relationship between the stellar velocity dispersion and star formation, SFR is plotted with velocity dispersion in Figure 6 for all blue ETGs in SCH09 for which SFR estimates are available. The upper limits for non-detections in WEL galaxies are also plotted here. It can be seen from Figure 6 that SFR in blue ETGs are linearly correlated with velocity dispersion up to nearly 100 km s$^{-1}$. At higher values of $\sigma$, SFR seems to be declining. It can be seen that for log $\sigma$ values between 2 and 2.2, significantly large fraction of galaxies shows less SFR compared to that expected from a linear relationship expected from low $\sigma$ values. A polynomial fit to all data points as shown in Figure 6 further establishes that SFR in blue ETGs declines beyond a velocity dispersion of nearly 100 km s$^{-1}$. The stellar velocity dispersion in galaxies is known to have strong links with the growth of black hole and triggering of AGNs, which in turn are believed to be regulating star formation \citep{1998A&A...331L...1S, 2012ARA&A..50..455F, 2006Natur.442..888S}. The black hole masses in centers of galaxies are found to be linearly correlated with stellar velocity dispersion \citep{1993nag..conf..197K, 1995ARA&A..33..581K, 1998AJ....115.2285M, 1998Natur.395A..14R, 2000ApJ...539L..13G, 2000ApJ...539L...9F, 2012Natur.491..729V, 2015MNRAS.446.2330S}. It is believed that AGN related shocks and winds remove the bulk of gas from galaxies thereby suppressing or halting star formation. Several studies have pointed out role of AGN feedback in regulating star formation in early-type galaxies \citep{2007MNRAS.382.1415S, 2011MNRAS.415.3798K, 2014MNRAS.440..889S, 2015MNRAS.452..774K}. A semi analytical study predicts that galaxies having stellar velocity dispersion in the range of 80 to 240 km s$^{-1}$ can experience suppression of star formation via the growth of black hole \citep{2006Natur.442..888S}. At higher velocity dispersion, the black hole merger dominates and star formation is almost prohibited in such galaxies. Our results for blue ETGs presented in this paper are in close agreement with these predictions. These results establishes that star formation in blue ETGs is predominantly found in galaxies with low stellar velocity dispersion, and is  suppressed in galaxies with higher velocity dispersion. This effect may in turn be linked to AGN feedback process in galaxies with high velocity dispersion. 

\begin{figure}
\includegraphics[width=7.5cm,height=9.0cm,angle=270]{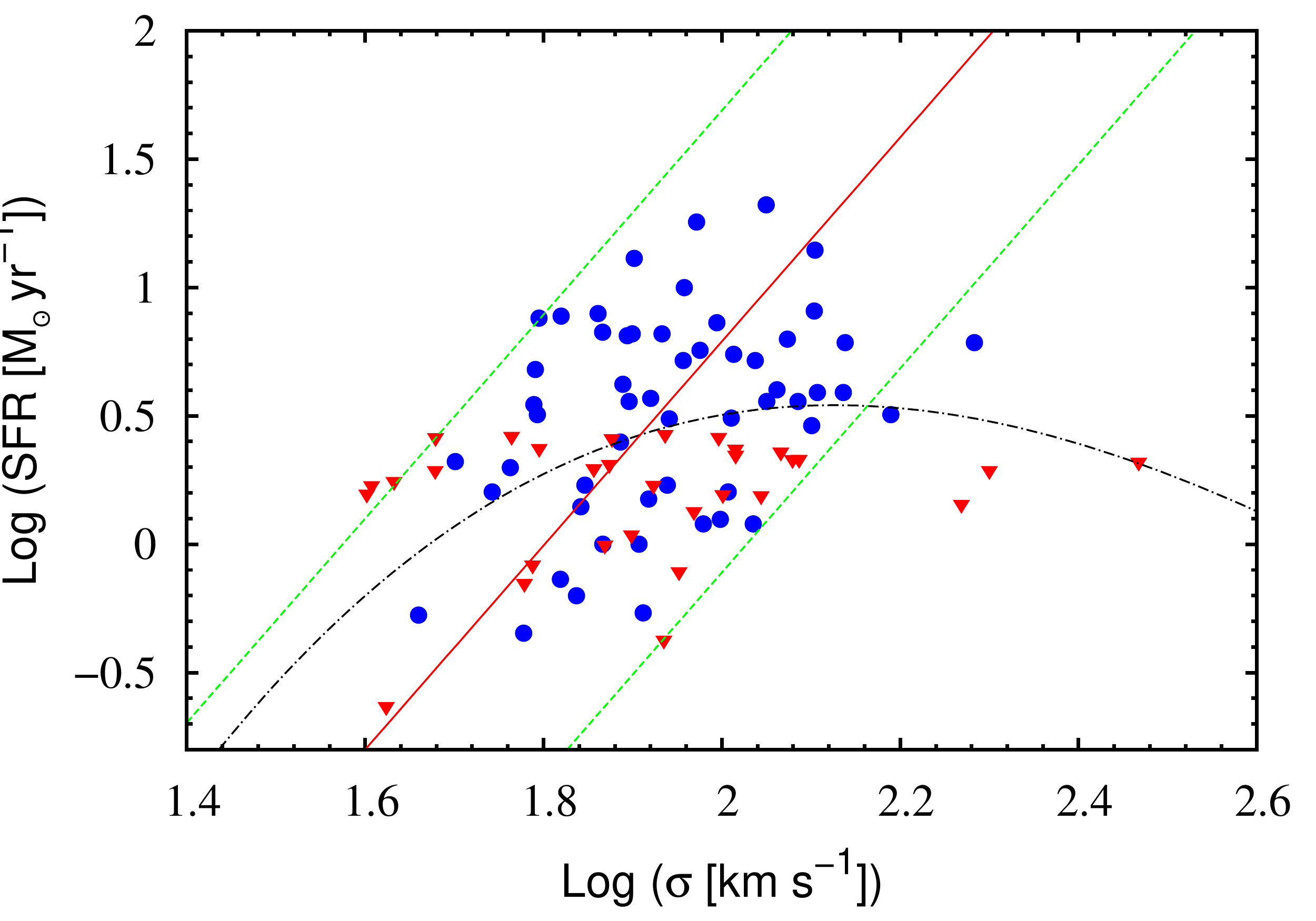}
\caption{The SFRs of blue ETGs are plotted against the stellar velocity dispersions. The SFR upper limits in WEL galaxies and the SFR in purely star forming galaxies are shown by triangle and circular points respectively. The solid line and dashed lines represent an average linear relation between SFR and $\sigma$ for $\sigma \textless 100$ km/s, and its limits to the scatter respectively. A polynomial fit (dot-dashed line) to all data points represents a possible trend of SFR with $\sigma$.}
\end{figure}

\begin{figure}
\centering
\includegraphics[width=9.5cm,height=8.0cm]{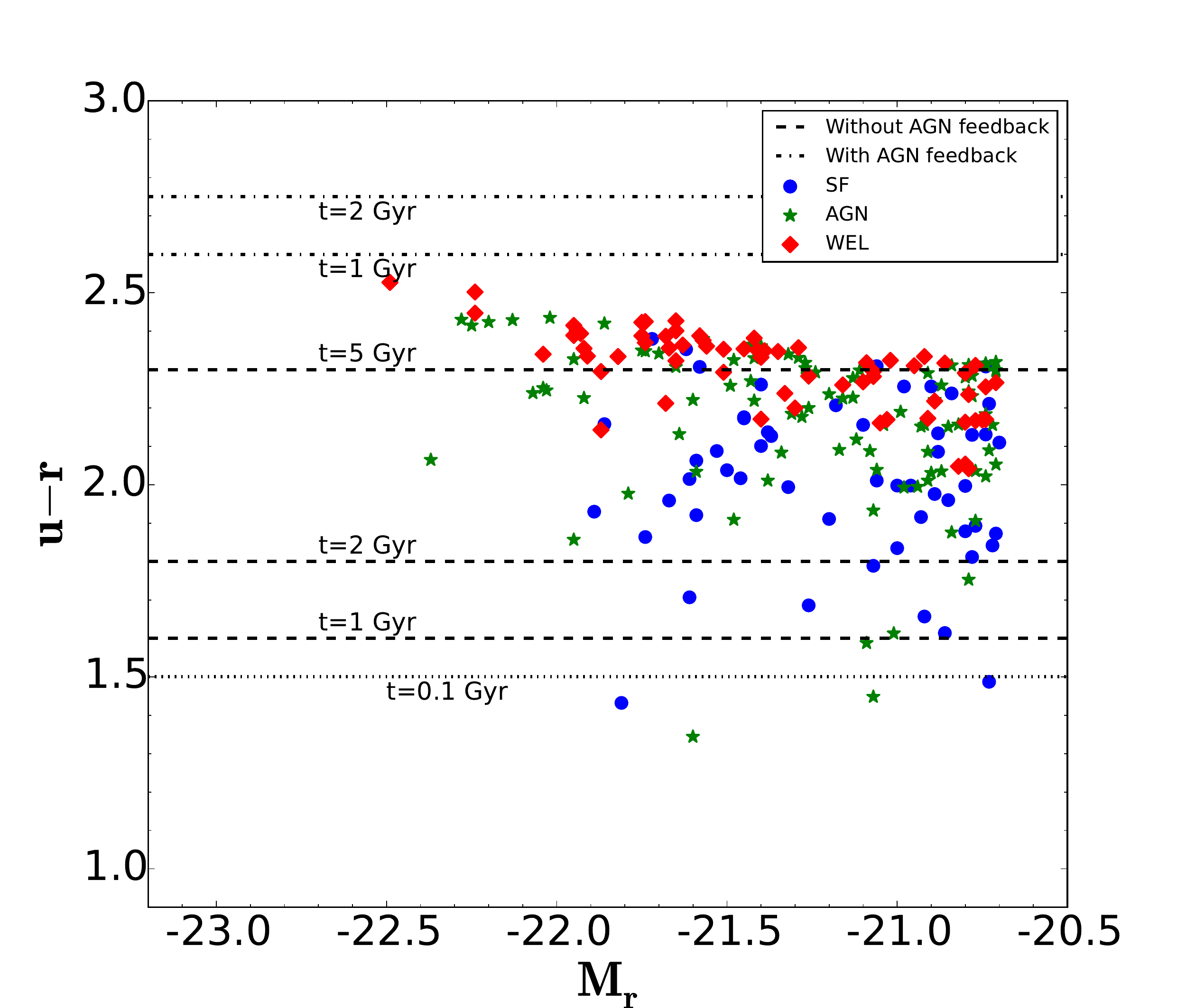}
\caption{The color distribution of all sub-types of blue ETGs in the present sample and their color evolution with time, fitted in two different scenarios - with and without AGN feedback.}
\end{figure}

In a study of 16,000 ETGs in the redshift range 0.05 - 0.1 taken from MOSES (Morphologically Selected Ellipticals in SDSS) data, \citet{2007MNRAS.382.1415S} identified an evolutionary sequence from SF to quiescence via AGN activities based on color differences. They predicted based on stellar evolution models that this transition process takes place within about 1 Gyr. SCH09 examined this scenario for the blue ETG sample. The color distribution of all blue ETGs is shown in Figure 7 along with time required for the color evolution based on two evolutionary models after an episode of starburst -- one assuming a passive evolution without AGN feedback and other with strong AGN feedback. These models are taken from \citet{2011MNRAS.415.3798K}. It can be seen here that the observed color distribution can be well fitted in terms of fast ($\sim 1$ Gyr) evolution through AGN feedback process. In absence of AGN feedback, the passive evolution will take very long time ($\textgreater$ 5 Gyr) to explain the observed color distribution. Based on this plot and the fact that a good fraction of blue ETGs have AGN activities, a causal relationship between AGN activities and star formation suppression can be inferred.

%%%%%%%%%%%%%%%%%%%%%%%%%%%%%%%%%%%%%%%%%%%%%%%%%%%%%%%%%
\section{Conclusions} %\label{sec:conclusions}
%%%%%%%%%%%%%%%%%%%%%%%%%%%%%%%%%%%%%%%%%%%%%%%%%%%%%%%%%

The 1.4 GHz radio continuum emission is detected directly in eight weak emission line galaxies with a radio luminosity in the range of $\sim$ 1 to 7 $\times$ 10$^{21}$ W Hz$^{-1}$. Their SFRs are in the range of $1 - 8$ M$_{\odot}$yr$^{-1}$. The statistically significant 1.4 GHz radio continuum emission of 79 $\mu$Jy with an rms of 19 $\mu$Jy beam$^{-1}$ is detected in a stacked image of 57 WEL galaxies. The median radio luminosity and SFR in these galaxies are estimated as 1.9 $\pm$ 0.5 $\times$ 10$^{20}$ W Hz$^{-1}$ and 0.23 $\pm$ 0.06 M$_{\odot}$yr$^{-1}$ respectively. The star formation rates in a volume-limited sample of pure star forming blue ETGs was known from earlier studies in the range of $0.5 - 50$ M$_{\odot}$yr$^{-1}$ based on strengths of optical emission lines from the SDSS. It was not possible earlier to estimate SFR in WEL galaxies due to weakness of optical line strengths in WEL galaxies. Based on radio continuum detections, we predicted SFR in majority of WEL galaxies at $\sim$ 0.2 M$_{\odot}$yr$^{-1}$. The radio far-infrared correlation was constructed in 4 WEL galaxies for which both radio and far-infrared detections were available. These galaxies follow the radio-FIR correlation without a significant deviation.

The SFRs in blue ETGs are found to be correlated with their stellar velocity dispersion. It was also shown here that average SFR in blue ETGs decreases gradually beyond $\sigma$ of $\sim 100$ km s$^{-1}$. This effect is most likely linked with the growth of black hole and suppression of star formation via AGN feedback process. The color differences between SF and WEL sub-types of blue ETGs appear to be driven to large extent by the level of current star formation activities within each sub-type. In the case of an evolutionary sequence between SF and WEL galaxies, the observed color distribution in blue ETGs can be explained best in terms of fast ($\sim 1$ Gyr) evolution through AGN feedback process. 
 
%%%%%%%%%%%%%%%%%%%%%%%%%%%%%%%%%%%%%%%%%%%%%%%%%%%%%%%%%
\section*{Acknowledgements}
%%%%%%%%%%%%%%%%%%%%%%%%%%%%%%%%%%%%%%%%%%%%%%%%%%%%%%%%%
We thank the referee for useful and constructive comments which greatly improved the content of the paper. We acknowledge the use of the VLA-FIRST survey and the IRAS survey. The FIRST sky survey was conducted by the National Radio Astronomical Observatory (NRAO) using the Very Large Array (VLA). The NRAO is a facility of the National Science Foundation operated under cooperative agreement by Associated Universities, Inc. The NASA/ IPAC Infrared Science Archive, which is operated by the Jet Propulsion Laboratory, California Institute of Technology, under contract with the National Aeronautics and Space Administration. This research has made use of the SAO/NASA Astrophysics Data System (ADS) operated by the Smithsonian Astrophysical Observatory (SAO) under a NASA grant. 

\bibliography{references}

\appendix

\label{lastpage}

\end{document}